\documentclass[showkeys,showpacs,superscriptaddress]{revtex4-2}
\usepackage{amsmath}
\usepackage{amssymb}
\usepackage{graphicx}
\usepackage{float}
\usepackage{xcolor}
\usepackage{physics}
\usepackage{ulem}

\allowdisplaybreaks

\begin{document}

\title{
Axially symmetric Proca-Higgs boson stars
}

\author{
Vladimir Dzhunushaliev
}
\email{v.dzhunushaliev@gmail.com}
\affiliation{
Department of Theoretical and Nuclear Physics,  Al-Farabi Kazakh National University, Almaty 050040, Kazakhstan
}
\affiliation{
Institute of Nuclear Physics, Almaty 050032, Kazakhstan
}
\affiliation{
Academician J.~Jeenbaev Institute of Physics of the NAS of the Kyrgyz Republic, 265 a, Chui Street, Bishkek 720071, Kyrgyzstan
}

\author{Vladimir Folomeev}
\email{vfolomeev@mail.ru}
\affiliation{
Institute of Nuclear Physics, Almaty 050032, Kazakhstan
}
\affiliation{
Academician J.~Jeenbaev Institute of Physics of the NAS of the Kyrgyz Republic, 265 a, Chui Street, Bishkek 720071, Kyrgyzstan
}
\affiliation{
International Laboratory for Theoretical Cosmology, Tomsk State University of Control Systems and Radioelectronics (TUSUR),
Tomsk 634050, Russia
}


\begin{abstract}
We consider strongly gravitating configurations consisting of coupled real Higgs scalar field and vector (Proca) field of mass $\mu_P$.
For such a system, we find static regular  axially symmetric solutions describing asymptotically flat configurations
which may be referred to as Proca-Higgs miniboson stars, since their total mass and spatial dimension are of order $M_{\text{Pl}}^2/\mu_P$ and $\mu_P^{-1}$, respectively.
The system possesses an axially symmetric dipole field and may be regarded as a Proca dipole.
\end{abstract}


\date{\today}

\maketitle

\section{Introduction}

Compact gravitating configurations supported by various fundamental fields have
been the object of vigorous investigations over the last decades.
This interest is partially a consequence of the fact that, according to the contemporary viewpoint,
such fields may play a crucial role both in describing the large-scale structure of the early and present Universe and
in modeling various clustered distributions of matter. In the first case, fundamental fields are employed, for example,
in modeling the inflationary stage in the early Universe and in describing its current accelerated expansion
(dark energy)~\cite{AmenTsu2010,Bamba:2012cp}. In turn, as applied to clustered distributions of matter, various fields are used both in modeling phenomena on
scales of galaxies and their clusters (dark matter) and in describing relatively small-scale configurations (for instance, stars).

In the latter case, there is prolific activity in studying compact gravitating objects supported by scalar
(spin-0) fields~-- the so-called boson stars. It is usually assumed that such configurations consist of different types of complex fields,
but the use of real scalar fields is also possible~\cite{Schunck:2003kk,Liebling:2012fv,Mielke:2016war}.
Depending on the type of the field, dimensions and masses of such stars may lie in a very wide range,
ranging from Planckian values to the scales comparable with characteristics that are typical for ordinary stars, or even for larger objects.

On the other hand, gravitating systems consisting of matter fields with nonzero spin are also under active study. In the case of half-integer-spin fields,
one may mention the spherically symmetric Einstein-Dirac systems consisting of both linear~\cite{Finster:1998ws,Herdeiro:2017fhv} and nonlinear spinor fields~\cite{Krechet:2014nda,Adanhounme:2012cm,Dzhunushaliev:2018jhj,Dzhunushaliev:2019kiy,Dzhunushaliev:2019uft}.
In the case of vector (spin-1) fields, considerable effort has been devoted to studying various
Einstein-Yang-Mills configurations pioneered in Ref.~\cite{Bartnik:1988am} (for a review, see, e.g., Ref.~\cite{Volkov:1998cc}).
In turn, in recent years interest in systems containing various massive vector (spin-1) fields~-- the so-called Proca stars~--
has increased considerably~\cite{Brito:2015pxa,Herdeiro:2017fhv,Minamitsuji:2018kof,Herdeiro:2019mbz,Bustillo:2020syj}.
Introduction of such fields is motivated by the fact that if one regards Proca theory as the generalization of Maxwell's theory,
it permits one to take into account various effects related
to the possible presence of the rest mass of a photon~\cite{Tu:2005ge},
to describe the massive $Z^0$ and $W^\pm$ particles in the Standard Model of particle physics~\cite{Lawrie2002}, to employ such
fields as applied to dark matter physics~\cite{Arkani-Hamed:2008hhe,Pospelov:2008jd}.

Another possible area of research is a consideration of gravitating systems involving several matter fields.
Investigations  of such systems can be arbitrarily divided into two directions.
First, it involves the consideration of mixed objects, which have one purely field component and another one is described using a variety of approximation schemes.
As an example, one can mention here (i)~systems consisting of bosonic and fermionic components, which either interact
only through a gravitational field~\cite{Henriques:1989ar} or also involve extra couplings~\cite{deSousa:1995ye}
(in both cases, the fermionic component is described by some effective equation of state), and
(ii)~configurations created by coupled fermion and scalar fields, when fermions are described using the Thomas-Fermi
approximation~\cite{Lee:1986tr}.
Within the second direction, purely field systems consisting of several components are under investigation.
These systems can be exemplified by (i)~spherically symmetric charged boson stars consisting of
a complex scalar field minimally coupled to an electric Maxwell field~\cite{Jetzer:1989av},
(ii)~spherically symmetric monopole solutions to the Einstein-Yang-Mills-Higgs equations~\cite{Breitenlohner:1991aa},
(iii)~axially symmetric monopole solutions supported by
Yang-Mills and dilaton~\cite{Kleihaus:1997mn} or Higgs~\cite{Kleihaus:2000hx,Hartmann:2001ic} scalar fields,
(iv)~spherically symmetric systems consisting of complex scalar and Proca fields~\cite{Brito:2015yfh}
or of spinor and electric Maxwell/Proca~\cite{Dzhunushaliev:2019kiy} or Yang-Mills/Proca~\cite{Dzhunushaliev:2019uft} fields.
Depending on the choice of concrete numerical values of the system parameters,
either the boson or fermion components can be the dominant part;
this ensures a wide range of physical properties of such configurations.

From what has been said above, it is of interest to continue studying the properties of configurations supported by several coupled
fields that are of different nature. In particular, this refers to gravitating systems created by non-Abelian fields coupled to scalar fields.
Here we consider a system consisting of a massive vector (Proca) field and a real Higgs field.
The use of the Proca field is interesting in the sense that its presence changes substantially the structure of the magnetic field
compared with, for example, that of the long-range magnetic field of the monopole solutions within Einstein-Yang-Mills-Higgs theory~\cite{Breitenlohner:1991aa,Kleihaus:2000hx,Hartmann:2001ic}.

As a starting configuration, we will use a nongravitating non-Abelian SU(2) Proca-Higgs system considered by us recently
in Ref.~\cite{Dzhunushaliev:2021oad}
(see also the related Refs.~\cite{Dzhunushaliev:2019ham,Dzhunushaliev:2019sxk,Dzhunushaliev:2020eqa,Dzhunushaliev:2021uit} where
various solutions for systems with coupled scalar, spinor, and vector fields have been studied).
At the microscopic level, such a system can be used in modeling localized field structures~-- particles/quasiparticles or tubes connecting them.
In the absence of gravity,
the system is in equilibrium due to the force balance: the non-Abelian field is purely repulsive, whereas the Higgs scalar field is purely attractive.
In turn, once the number of particles becomes large, it may already become necessary to take into account the gravitational interaction between the particles,
and this leads to the appearance of the additional attractive forces. Our task here is to examine the properties of such a system containing a large number of particles
minimally coupled to gravity.

In the absence of gravity,
the system of Ref.~\cite{Dzhunushaliev:2021oad} can contain both electric and magnetic fields simultaneously.
However, when a gravitational field is involved,
in SU(2) theory, four-dimensional configurations with an asymptotically flat spacetime
can be only purely magnetic~\cite{Ershov:1990qwn}. Consistent with this, we will consider here a case where only
a magnetic field is present.
In this case, the  SU(2) system of Ref.~\cite{Dzhunushaliev:2021oad} reduces to
an  embedded Abelian U(1) system with minimally coupled magnetic and scalar fields. Our purpose will be to study the dependence
of the properties of such a system on the value of the coupling constant.

The paper is organized as follows. In Sec.~\ref{Proca_Dirac_scalar}, we write down the general field equations for the non-Abelian Einstein-Proca-Higgs theory.
In Sec.~\ref{axi_sol}, we solve these equations numerically and
obtain axially symmetric solutions describing compact gravitating configurations consisting of a magnetic Proca field coupled to a Higgs scalar field.
Finally, Sec.~\ref{concl} summarizes the results obtained in the paper.

\section{Einstein-Proca-Higgs theory}
\label{Proca_Dirac_scalar}

Consistent with the purpose of the study given in the Introduction, let us generalize the system considered in
 Refs.~\cite{Dzhunushaliev:2021oad,Dzhunushaliev:2019sxk,Dzhunushaliev:2021uit} to the case of the presence of
a strong gravitational field. In this case, the Lagrangian describing a system consisting of a non-Abelian SU(3)
Proca field $A^a_\mu$ interacting with nonlinear scalar field $\phi$ can be taken in the form
[hereafter, we work in units such that $c=\hbar=1$ and the metric signature is $(+, -, -, -)$]
\begin{equation}
	\mathcal L = -\frac{R}{16\pi G}
 - \frac{1}{4} F^a_{\mu \nu} F^{a \mu \nu} -
	\frac{1}{2}\mathfrak{m}^{a b, \mu}_{\phantom{a b,}\nu}
	A^a_\mu A^{b \nu} +
	\frac{1}{2} \partial_\mu \phi \partial^\mu \phi +
	\frac{\lambda}{2} \phi^2 A^a_\mu A^{a \mu} -
	\frac{\Lambda}{4} \left( \phi^2 - M^2 \right)^2.
\label{0_10}
\end{equation}
Here  $G$ is the Newtonian gravitational constant, $R$ is the
scalar curvature,
$
	F^a_{\mu \nu} = \partial_\mu A^a_\nu - \partial_\nu A^a_\mu +
	g f_{a b c} A^b_\mu A^c_\nu
$ is the field strength tensor for the Proca field, where $f_{a b c}$ are the SU(3) structure constants, $g$ is the coupling constant,
$a,b,c = 1,2, \dots, 8$ are color indices, and
$\mu, \nu = 0, 1, 2, 3$ are spacetime indices. The Lagrangian \eqref{0_10} also contains the arbitrary constants $M, \lambda$, and $\Lambda$ and the Proca field mass matrix
$
	\mathfrak{m}^{a b, \mu}_{\phantom{a b,}\nu}
$.

Making use of the Lagrangian~\eqref{0_10}, the corresponding field equations can be written in the form
\begin{align}
R_{\mu}^\nu - \frac{1}{2} \delta_{\mu }^\nu R =&
	8\pi G\, T_{\mu }^\nu,
\label{feqs-10}\\
	\frac{1}{\sqrt{-\mathcal G}} \frac{\partial}{\partial x^\nu}\left(\sqrt{-\mathcal G}F^{a \mu \nu}\right) + g f_{a b c} A^b_\nu F^{c \mu \nu}
 =& \lambda \phi^2 A^{a \mu}
	- \mathfrak{m}^{a b, \mu}_{\phantom{a b,}\nu} A^{b \nu},
\label{0_20}\\
	\frac{1}{\sqrt{-\mathcal G}} \frac{\partial}{\partial x^\mu} \left(\sqrt{-\mathcal G}\mathcal G^{\mu\nu}\frac{\partial\phi}{\partial x^\nu} \right)
=& \lambda A^a_\mu A^{a \mu} \phi +
	\Lambda \phi \left( M^2 - \phi^2 \right) ,
\label{0_30}
\end{align}
where $\mathcal G_{\mu\nu}$ is the spacetime metric. In turn,
the right-hand side of Eq.~\eqref{feqs-10} contains the energy-momentum tensor
\begin{align}
\label{EM}
\begin{split}
	T_{\mu}^\nu =
&\mathcal G^{\nu\sigma}\partial_\sigma \phi\partial_\mu \phi-\delta^\nu_\mu\left[
\frac{1}{2}\mathcal G^{\lambda\sigma}\partial_\sigma \phi\partial_\lambda \phi-\frac{\Lambda}{4} \left( \phi^2 - M^2 \right)^2
\right]-F^{a\nu\rho}F^a_{\mu\rho}+\frac{1}{4}\delta_\mu^\nu F^a_{\alpha\beta}F^{a\alpha\beta}\\
&
 -\mathfrak{m}^{a b, \alpha}_{\phantom{a b,}\mu}  A^a_\alpha A^{b \nu}+
\frac{1}{2}\delta^\nu_\mu \mathfrak{m}^{a b, \alpha}_{\phantom{a b,}\beta} A^a_\alpha A^{b \beta}+
\lambda \phi^2\left(A^{a\nu}A_{a\mu}-\frac{1}{2}\delta^\nu_\mu A^{a\alpha}A_{a\alpha}
\right).
\end{split}
\end{align}

\section{Axially symmetric solutions}
\label{axi_sol}

As pointed out in the Introduction, we will consider a particular case where the system contains only a magnetic Proca field and a scalar field.
In doing so, as in Ref.~\cite{Dzhunushaliev:2021oad},  we assume for simplicity that there is only one nonzero
component of the vector potential, $A^7_\varphi\neq 0$, describing the magnetic field.
This, in turn, implies that we will deal with an axially symmetric problem.

\subsection{The ansatz and equations}

We will study static axially symmetric configurations, for which it is convenient to choose the spacetime metric in the Lewis-Papapetrou form
\begin{equation}
	ds^2 = f dt^2-\frac{m}{f}\left(d r^2+r^2 d\theta^2\right)-\frac{l}{f}r^2\sin^2\theta d\varphi^2,
\label{metric}
\end{equation}
where the metric functions $f, l$, and $m$ depend on $r$ and $\theta$ only. The $z$ axis ($\theta=0$) represents the symmetry axis of the system.
Asymptotically (as $r\to \infty$), the functions $f, m, l \to 1$; i.e., the spacetime approaches a flat, Minkowski spacetime.

For the Proca field, we take here a purely magnetic ansatz (cf. Ref.~\cite{Dzhunushaliev:2021oad})
\begin{equation}
			A^7_\varphi = r\sin{\theta}\frac{w(r, \theta)}{g} .
\label{vec_pot}
\end{equation}
In this case, the general system~\eqref{0_10} reduces, in essence, to an embedded Abelian U(1) system, where U(1)~$\subset$~SU(3).

For the above ansatz, there are the following nonzero physical components of the magnetic field:
\begin{equation}
	H^7_r =   -\frac{f}{\sqrt{l m}}\frac{w_{, \theta} + w \cot \theta}{g r } , \quad
	H^7_\theta =\frac{f}{\sqrt{l m}} \frac{w + r w_{,r}}{g r} .
\label{2_20}
\end{equation}
(Henceforth a comma in lower indices denotes differentiation with respect to the corresponding coordinate.)

Since we consider here only one component of the vector field~\eqref{vec_pot}, the Proca field mass matrix contains only the component
$\mu_P^2\equiv\mathfrak{m}^{7 7, \varphi}_{\phantom{a b,}\varphi}$.
Then, introducing the dimensionless variables
\begin{equation}
x=\mu_P r, \quad \bar\phi, \bar w, \bar M=\sqrt{8\pi G}\{\phi, w, M\}, \quad \bar \lambda, \bar \Lambda=\frac{\{\lambda, \Lambda\}}{8\pi G \mu_P^2},
\label{dmls_var}
\end{equation}
the field equations \eqref{feqs-10}-\eqref{0_30}, when the expression~\eqref{EM} is inserted, yield
\begin{align}
	 f_{,xx} + \frac{f_{,\theta \theta}}{x^2}
	+ \frac{2f_{,x}}{x} + \frac{\cot \theta f_{,\theta}}{x^2}
	- \frac{1}{f}
	\bigg(
		f_{,x}^2 + \frac{f_{,\theta}^2}{x^2}
	\bigg)
	+ \frac{1}{2l}
		\bigg(f_{,x}l_{,x}+\frac{f_{,\theta}l_{,\theta}}{x^2}
	\bigg) &
\nonumber\\
	- \frac{f^2}{g^2 x^2 l}
	\left[
		2w\left(
			\cot\theta w_{,\theta} + x w_{,x}
		\right)
		+ w_{,\theta}^2 + x^2 w_{,x}^2 + \csc^2\theta w^2
	\right]
	+ \frac{\Lambda}{2} m \left(\phi^2 - M^2\right)^2
	= & 0,
\label{2_30}\\
	 l_{,xx} + \frac{l_{,\theta \theta}}{x^2}
	+ \frac{3l_{,x}}{x} + \frac{2\cot \theta l_{,\theta}}{x^2}
	- \frac{1}{2l} \bigg(
		l_{,x}^2+\frac{l_{,\theta}^2}{x^2}
	\bigg)
	+ \Lambda\frac{l m}{f}
	\left(\phi^2-M^2\right)^2 +
	\frac{2 m}{g^2} \left(\lambda\phi^2 - 1\right)w^2
	= & 0,
\label{2_40}\\
	m_{,xx} + \frac{m_{,\theta \theta}}{x^2}
	+ \frac{m_{,x}}{x}
	+ \frac{m}{2 f^2}
	\bigg(
		f_{,x}^2+\frac{f_{,\theta}^2}{x^2}
	\bigg)
	- \frac{1}{m}
	\bigg(
		m_{,x}^2 + \frac{m_{,\theta}^2}{x^2}
	\bigg) +
	m \left(
		\phi_{,x}^2 + \frac{\phi_{,\theta}^2}{x^2}
	\right)
	+ \frac{\Lambda}{2}\frac{m^2}{f}\left(\phi^2-M^2\right)^2 &
\nonumber\\
	- \frac{m}{g^2 x^2 l}
	\left\{
		\left[
		\csc^2\theta f + x^2 m\left(\lambda\phi^2-1\right)
		\right] w^2	
		+ f\left[
			w_{,\theta}^2 + x^2 w_{,x}^2 + 2 w
				\left(\cot\theta w_{,\theta}+x w_{,x}\right)
		\right]
	\right\} = & 0,
\label{2_50}\\
	\phi_{,xx} + \frac{\phi_{,\theta \theta}}{x^2}
	+ \left(
	\frac{2}{x} + \frac{ l_{,x}}{2l}
	\right) \phi_{,x}
	+ \left(\cot \theta + \frac{ l_{,\theta}}{2l}\right)  \frac{\phi_{,\theta}}{x^2}
	+ \left[
		\Lambda\frac{m}{f}\left(M^2 - \phi^2\right)
		- \frac{\lambda}{g^2} \frac{m}{l}w^2
	\right] \phi = & 0,
\label{2_60}\\
	w_{,xx} + \frac{w_{,\theta \theta}}{x^2}
	+ \left(
		\frac{2}{x} + \frac{f_{,x}}{f} - \frac{l_{,x}}{2 l}
	\right) w_{,x} + \frac{\cot \theta w_{,\theta}}{x^2}
	- \frac{1}{2x^2 f l}
	\left(f l_{,\theta} - 2 l f_{,\theta}\right)
	w_{,\theta} &
\nonumber\\
	+ \frac{1}{x^2}
	\left(
		\frac{\cot\theta f_{,\theta} + x f_{,x}}{f}
		- \frac{2\csc^2\theta\, l + \cot\theta\, l_{,\theta}
		+ x l_{,x}}{2l}
	\right)w + \frac{m}{f} \left(1 - \lambda \phi^2\right) w = &0.
\label{2_70}
\end{align}
Here Eqs.~\eqref{2_30}-\eqref{2_50} are combinations of the components of the Einstein equations $E_t^t-E_r^r-E_\theta^\theta-E_\varphi^\varphi=0$,
$E_r^r+E_\theta^\theta=0$, and $E_\varphi^\varphi=0$, respectively.
To make the notation simpler, we have omitted in these equations the bar sign over the dimensionless variables.

\subsection{Boundary conditions}

We will seek globally regular, asymptotically flat solutions possessing a finite mass. For such solutions, we impose
appropriate boundary conditions for the Proca/scalar fields and metric functions at the origin ($x=0$), at infinity ($x\to \infty$),
on the positive $z$ axis ($\theta=0$), and, making use of the reflection symmetry with respect to $\theta\to \pi-\theta$,
in the equatorial plane ($\theta=\pi/2$). So we require
\begin{align}
&\left. \frac{\partial f}{\partial x}\right|_{x = 0} =
\left. \frac{\partial m}{\partial x}\right|_{x = 0} =
\left. \frac{\partial l}{\partial x}\right|_{x = 0} =
	\left. \frac{\partial \phi}{\partial x}\right|_{x = 0} =  0,  \left. w \right|_{x = 0} = 0;\nonumber\\
&\left. f \right|_{x = \infty} = \left. m \right|_{x = \infty} =\left. l \right|_{x = \infty} =1 ,
	\left. w \right|_{x = \infty} = 0 ,
	\left. \phi \right|_{x = \infty} =  M ; \nonumber\\
&\left. \frac{\partial f}{\partial \theta}\right|_{\theta = 0,\pi} =\left. \frac{\partial m}{\partial \theta}\right|_{\theta = 0,\pi} =
\left. \frac{\partial l}{\partial \theta}\right|_{\theta = 0,\pi} =
	\left. \frac{\partial \phi}{\partial \theta}\right|_{\theta = 0,\pi} =  0 ,  \left. w \right|_{\theta = 0,\pi} = 0 ;\nonumber\\
&\left. \frac{\partial f}{\partial \theta}\right|_{\theta = \pi/2} =\left. \frac{\partial m}{\partial \theta}\right|_{\theta = \pi/2} =
\left. \frac{\partial l}{\partial \theta}\right|_{\theta = \pi/2} =
	\left. \frac{\partial w}{\partial \theta}\right|_{\theta = \pi/2} =
	\left. \frac{\partial \phi}{\partial \theta}\right|_{\theta = \pi/2} =  0.\nonumber
\end{align}
In turn, the condition of the absence of a conical singularity requires that the solutions should satisfy the
constraint $\left. m\right|_{\theta=0,\pi}=\left. l\right|_{\theta=0,\pi}$ (we have been checking the fulfilment of these conditions in performing calculations).

\subsection{Asymptotic behavior}

Before proceeding to the discussion of the solutions, let us write down the expressions describing an asymptotic behavior of the matter fields.
It is assumed that, as $x\to \infty$, the spacetime approaches Minkowski spacetime; i.e., the metric functions $f,m,$ and $l$ approach unity.
In this case, since we seek here asymptotically decaying solutions for which $w\to 0$ exponentially fast,
one can neglect the nonlinear term in Eq.~\eqref{2_60} containing  $w^2$. In turn, asymptotically, the field
$
	\phi \approx M - \eta \rightarrow M
$, and the function $\eta$ decays exponentially; this permits us to replace the term $\phi^2$ by $M^2$ in Eq.~\eqref{2_70}.
As a result, from Eqs.~\eqref{2_60} and \eqref{2_70} , one can derive the following asymptotic equations:
\begin{align}
  \bigtriangleup_{x, \theta} w - \frac{w}{x^2\sin^2{\theta}}  +
	\left(1 -  \lambda M^2\right) w =& 0 ,
\nonumber\\
	  \bigtriangleup_{x, \theta} \eta -
	2  \Lambda M^2 \eta =& 0 ,
\nonumber
\end{align}
where $\bigtriangleup_{x, \theta}$ is the Laplacian operator in flat space. These equations have obvious solutions of the form
\begin{align}
	w \approx & C_{w}F(\theta)
		\frac{e^{- x \sqrt{\lambda M^2 - 1}}}{x},
\label{2_120}\\
	\eta \approx & C_{\eta}
	\left( Y\right)^0_{l_\eta}
	\frac{e^{- x \sqrt{2 \Lambda M^2}}}{x},
\label{2_130}
\end{align}
where $\left( Y\right)^0_{l_{\eta}}$ is a spherical function and $C_{w, \eta}$ are constants. In turn, the angular part of Eq.~\eqref{2_120}
is expressed in terms of special functions collected in $F(\theta)$
(we do not show this expression here to avoid overburdening the text).
It follows from Eq.~\eqref{2_120} that there is a lower limit on the combination of the parameters $M$ and $\lambda$ ensuring
the exponential asymptotic decay of the solution: $\lambda M^2>1$.

\subsection{Numerical solutions}

The set of five coupled nonlinear elliptic partial differential equations~\eqref{2_30}-\eqref{2_70}  has been solved numerically subject to the above boundary conditions.
 For numerical calculations, it is convenient to introduce a new compactified radial coordinate
\begin{equation}
\bar x=\frac{x}{1+x},
\label{comp_coord}
\end{equation}
the use of which permits one to map the infinite region $[0,\infty)$ to the finite interval $[0,1]$. Calculations have been carried out using the package FIDISOL~\cite{fidisol}
with typical errors on the order of $10^{-4}$. In turn, in plotting graphs, we have used compactified coordinates
\begin{equation}
\bar\rho=\frac{\bar x \sin\theta}{1-\bar x\left(1-\sin\theta\right)}, \quad
\bar z=\frac{\bar x \cos\theta}{1-\bar x\left(1-\cos\theta\right)},
\label{comp_coord_cyl}
\end{equation}
where $\bar x$ is given by Eq.~\eqref{comp_coord}; these coordinates cover all space of the solutions.

\begin{figure}[h!]
\includegraphics[width=.99\linewidth]{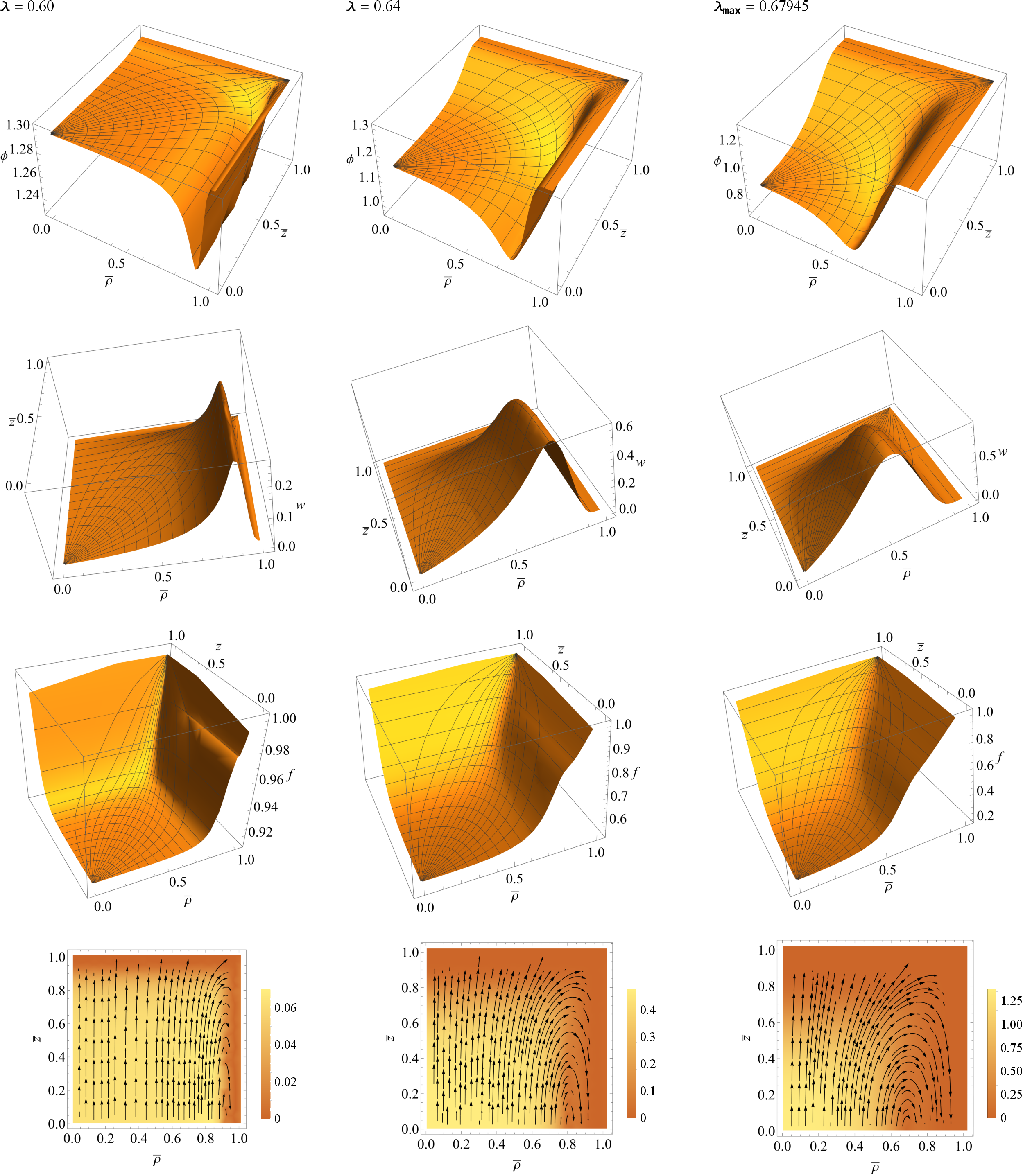}
\caption{
	Distributions of the dimensionless fields $\phi$ (upper row) and $w$ (second row), the metric function	$f$ (third row),
and the strength of the magnetic field $\vec H^7$  (expressed in units $\mu_P/\sqrt{8\pi G}$) from Eq.~\eqref{2_20} (lower row)	
for different values of $\lambda$ and fixed values of the parameters $M=1.3$, $\Lambda=0.2$, and $g=1$.	
The plots for the magnetic field strength are made in a meridional plane $\varphi = \text{const.}$ spanned by the coordinates~\eqref{comp_coord_cyl}.
Since the system is mirror symmetric with respect to the equatorial plane $\bar z=0$,
we show only the solutions lying in the upper hemisphere $\bar z>0$.
}
\label{fig_fields_distr}
\end{figure}

As an example of localized solution, Fig.~\ref{fig_fields_distr} shows the distributions of the matter fields $\phi, w$ and
the metric function $f$ for fixed values of $M$ and $\Lambda$ and three different values of the coupling constant $\lambda$.
For the values of $M$ and $\Lambda$ given in the caption of the figure, the value $\lambda=0.6$
is close to the minimum permissible value $\lambda_{\text{min}}$ for which the asymptotically decaying solutions~\eqref{2_120} and \eqref{2_130}
are still valid;  for the solutions given in Fig.~\ref{fig_fields_distr}, the value $\lambda_{\text{min}}= 1/M^2\approx 0.592$.

Technically, in order to obtain solutions with different $\lambda$, we have used the following step-by-step procedure: First, we find a solution
at the step $n$ for some $\lambda_n>\lambda_{\text{min}}$ to an accuracy of the order of $10^{-4}$.
Then this solution is used as an initial guess for finding a solution at the step $n+1$, where
$\lambda_{n+1}=\lambda_n+\delta \lambda$ with $\delta \lambda \ll \lambda_n$. Proceeding in this manner,
we are eventually able to find a solution for some value $\lambda=\lambda_{\text{max}}$ for which a further increase of
$\lambda$ requires a considerable decrease of $\delta \lambda$ to keep the required accuracy.
Consistent with this, we have been terminating the calculations for some judicious values of $\delta \lambda \to 0$.
The resulting solutions correspond to the case of $\lambda=\lambda_{\text{max}}$, and they are exemplified
in the right column of Fig.~\ref{fig_fields_distr} for the case of $\Lambda=0.2$.
A similar situation occurs for other values of the parameter  $\Lambda$ as well (see below).


It is seen from the structure of the magnetic field strength
depicted in the lower row of Fig.~\ref{fig_fields_distr} that the system possesses
 an axially symmetric dipole field sourced by the current
associated with the scalar field and given by the first term on the right-hand side of Eq.~\eqref{0_20}.
This current is located in the equatorial plane of the configuration; its location along the radius and the magnitude
of the magnetic field are determined by the value of the parameter~$\lambda$:
as~$\lambda$ increases, the location of the current shifts to the central region and the field strength grows.
In turn,
with decreasing~$\lambda$, the distributions of all the fields become
more uniform, and for $\lambda\approx \lambda_{\text{min}}$ the configuration already possesses a sufficiently weak gravitational field
(the metric function $f$ is rather close to unity) and a practically uniformly distributed weak magnetic field (see the leftmost panel in the lower row of Fig.~\ref{fig_fields_distr}).
In this connection, one may expect that, as
$\lambda\to \lambda_{\text{min}}$, the system will tend to its vacuum state with $\phi=M$ and $w=0$.

\begin{figure}[t]
\includegraphics[width=.5\linewidth]{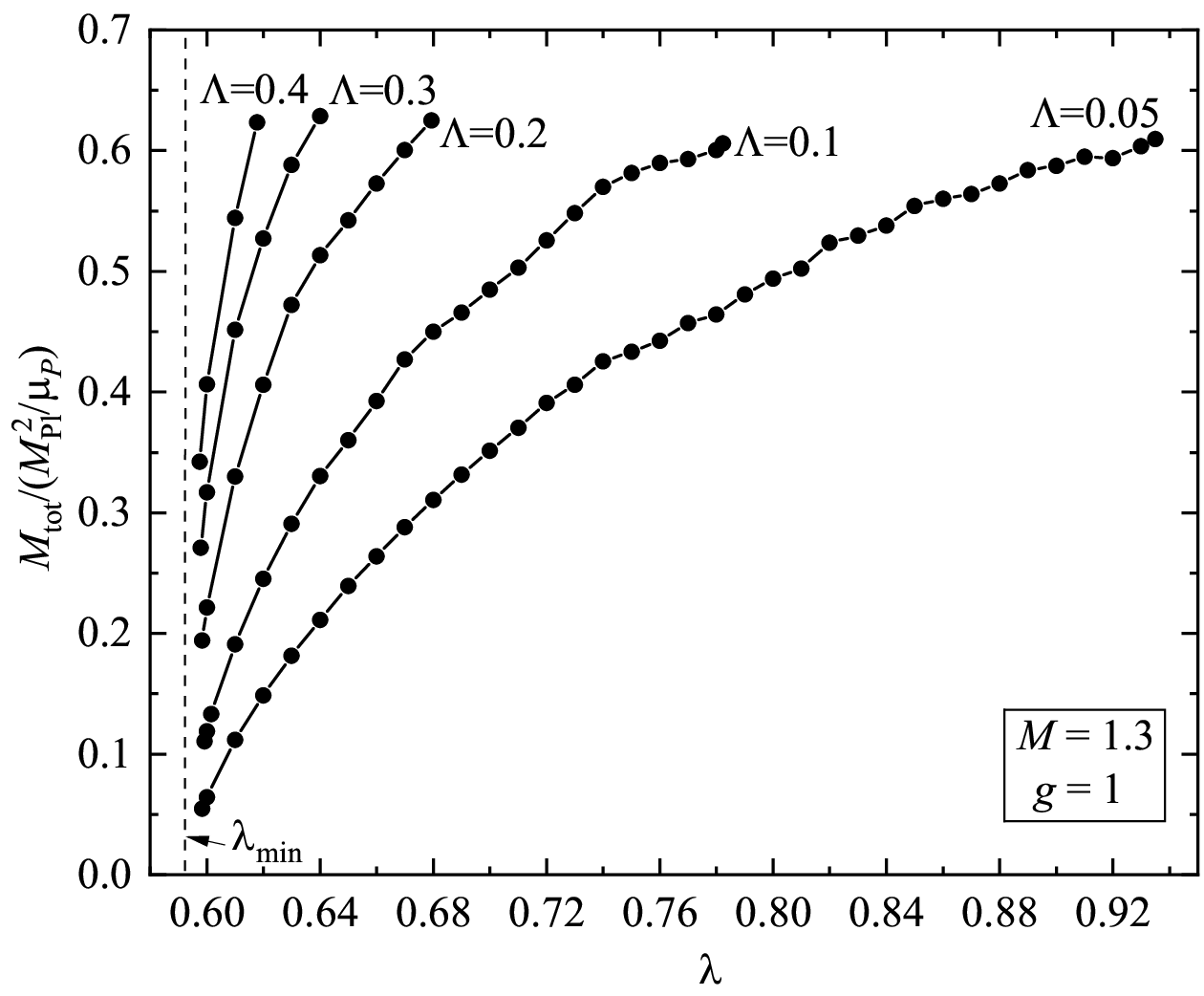}
\vspace{-0.5cm}
\caption{
	The dependence of the total mass \eqref{M_tot} of the configurations under consideration on $\lambda$ for different $\Lambda$.
}
\label{fig_mass_tot}
\end{figure}

This may be more clearly demonstrated by considering the behavior of the total mass of the objects under investigation for different values of the system parameters.
The static configurations under consideration are topologically trivial and globally regular (without an event horizon or conical singularities).
Bearing in mind that the spacetime of such systems is asymptotically flat, one can calculate their total mass using
a definition of the mass via the Komar integral~\cite{Wald}
$$
M_{\text{tot}}=2\int_{\Sigma} \left(T_{\mu \nu}-\frac{1}{2}\mathcal G_{\mu \nu}T\right)n^\mu \xi^\nu dV,
$$
where $n^\mu$ is a normal to $\Sigma$ and
$\xi^\nu$ is a timelike Killing vector.
Making use of the dimensionless variables~\eqref{dmls_var}, the metric~\eqref{metric}, and the energy-momentum tensor~\eqref{EM},
one can find 
\begin{equation}
	M_{\text{tot}} =
	\frac{1}{4 g^2} \frac{M_{\text{Pl}}^2}{\mu_P}
	\int_0^\infty dx \int_0^\pi d\theta\,
	\frac{\sin\theta}{f\sqrt{l}}
	\left\{
	f^2\left[
		w_{,\theta}^2 + x^2 w_{,x}^2 + 2 w
		\left(
			\cot\theta w_{,\theta} + x w_{,x}
		\right) + \csc^2\theta \,w^2
	\right]
	- \frac{\Lambda}{2}g^2 x^2 l m \left(\phi^2-M^2\right)^2
	\right\},
\label{M_tot}
\end{equation}
where $M_{\text{Pl}}$ is the Planck mass. Alternatively, the total mass of the configuration can be read off from the asymptotic expansion of the metric function
$
\left. f \right|_{r \to \infty}\approx 1-2 G M_{\text{tot}}/r.
$
Both these ways of calculating the mass can be employed to control the correctness of computations.

The results of calculations for the total mass are given in Fig.~\ref{fig_mass_tot}. These graphs are plotted for all the range of values of~$\lambda$
for which we have succeeded in obtaining numerical solutions to the required accuracy. It is seen from the figure that, as $\lambda$ approaches $\lambda_{\text{min}}$,
the total mass becomes an increasingly rapidly decaying function of $\lambda$ for all $\Lambda$, and one may expect that as $\lambda\to\lambda_{\text{min}}$,
the mass $M_{\text{tot}}$ will approach zero. On the other hand, with increasing $\lambda$, the total mass grows, reaching the value $M_{\text{tot}}\approx 0.61-0.62$
for all $\Lambda$ for some $\lambda_{\text{max}}=\lambda_{\text{max}}(\Lambda)$. However, it is seen from the behavior of the graphs that, at this limiting point,
$M_{\text{tot}}$ still does not approach saturation, and if only the numerical technique could permit us to find a solution for
$\lambda>\lambda_{\text{max}}$, one might expect that the total mass would continue its growth as $\lambda$ increases.
This is also indirectly confirmed by the fact that the metric function $f$, as well as two other metric functions $m$  and $l$,
remain finite and differ substantially from zero for the configurations with maximum values of
$\lambda$ for which we have succeeded in performing calculations (cf. Fig.~\ref{fig_fields_distr},
which shows the distribution of the function $f$ in the case of $\Lambda=0.2$ for which $\lambda_{\text{max}}=0.67945$).

\section{Conclusions}
\label{concl}

The purpose of the present paper is to study axisymmetric strongly gravitating configurations supported by
interacting real Proca and Higgs fields. In studying Proca fields, both in gravity and in a flat spacetime,
a natural question arises as to the existence of such fields in nature. As we pointed out in Ref.~\cite{Dzhunushaliev:2021oad},
there could be two possible answers to this question. The first one suggests that Proca fields are fundamental, and they do exist in nature.
This implies the violation of the concept that all fundamental integer-spin fields must be gauge invariant.
The second point of view is that Proca fields are phenomenological, and they arise as a result of some approximate description of other fundamental fields.
For example, this can happen in quantizing some fundamental field when the corresponding quanta of such field acquire an effective mass;
here one may draw an analogy to spontaneous symmetry breaking when some quanta of a gauge field acquire a mass.

Unlike the nongravitating configurations in SU(2) theory studied by us in Ref.~\cite{Dzhunushaliev:2021oad},
the presence of a gravitational field does not permit one to obtain localized, asymptotically flat systems involving both electric and magnetic fields.
For this reason, we have considered here the case with a magnetic field only, limiting ourselves for simplicity to
static systems containing only one component of the vector potential $A^7_\varphi$.
For such a case, we have found localized regular asymptotically flat solutions describing configurations
with the total mass~$\sim M_{\text{Pl}}^2/\mu_P$ and a spatial dimension~$\sim \mu_P^{-1}$.
Such objects may be referred to as ``Proca-Higgs miniboson stars,'' since
their masses and sizes are comparable to the characteristics of miniboson stars supported by a complex scalar field~\cite{Kaup:1968zz,Friedberg:1986tp}.
An essential point here is that if in the case of stars with a complex scalar field their characteristics are determined by the mass of the scalar field,
in our case this is the mass of the vector field~$\mu_P$.

Notice the following features of the configurations obtained:
\begin{itemize}
\item When considering gravitating configurations supported by real scalar fields only, there are no nonsingular,
static solutions with a trivial spacetime topology~\cite{Schunck:2003kk,Friedberg:1986tp,Jetzer:1992np}.
Aside from this, to the best of our knowledge, even in considering a complex Higgs scalar field, there are no regular, starlike solutions as well.
Regular gravitating solutions are only possible in the case of a nontrivial spacetime topology
when there are topological (kinklike) solutions for a real ghost Higgs scalar field~\cite{Kodama:1978dw,Kodama:1978zg}.
We have demonstrated here that in the presence of a
coupling between real Higgs and Proca fields the existence of static localized regular solutions is possible.
In this case, the spacetime topology is trivial and the solutions for the scalar field are nontopological.
\item On comparing the configurations obtained here and miniboson stars supported by a complex scalar field,
an important difference is that the latter systems contain a free
eigenparameter~-- the boson frequency, whose presence permits one to get configurations possessing different central densities
for fixed values of other field parameters~\cite{Schunck:2003kk,Liebling:2012fv}.
The magnitude of the central density determines the total mass of the system and permits one to
estimate, for example, the stability of such objects. On the contrary, in the case of a real scalar field considered in the present paper, there is no such a free
eigenparameter, and for fixed values of the scalar-field parameters ($\Lambda$ and $M$) and of the coupling constant $\lambda$,
there is only one (eigen) value of the scalar-field central density, and, correspondingly, of the total mass of the system.
\item The numerical calculations indicate that, depending on the specific values of the scalar-field parameters and
of the coupling constant $\lambda$, the total masses of the configurations lie in the range from zero to some finite value,
which turns out to be approximately identical for different values of $\Lambda$ (see Fig.~\ref{fig_mass_tot}). Unfortunately, further
calculations are limited by the numerical accuracy.
\item
The system possesses an axially symmetric dipole field sourced by the current associated with the Higgs field.
In this connection, such a configuration may be regarded as a ``Proca dipole,'' whose magnetic field strength
grows with increasing $\lambda$. In turn, asymptotically,
this field decreases exponentially with distance, in contrast to a monopole magnetic field (both in the spherically~\cite{Breitenlohner:1991aa}
and axially~\cite{Kleihaus:2000hx,Hartmann:2001ic} symmetric cases)
or a dipole field in Maxwell's electrodynamics, which decrease away from the source according to a power law.
\end{itemize}

In conclusion, let us briefly address the question of stability of the systems under investigation.
In the case of an asymptotically flat spacetime,
all static, spherically symmetric and purely magnetic regular or black hole solutions to the Einstein-Yang-Mills
equations for arbitrary gauge groups are known to be unstable~\cite{Volkov:1998cc}. In turn,
configurations supported only by a real scalar field are also dynamically unstable, whether they consist of ordinary~\cite{Jetzer:1992np} or ghost~\cite{Dzhunushaliev:2008bq} scalar fields
(see, however, Ref.~\cite{Clayton:1998zza} where the stability analysis for the ordinary scalar field was revisited and the possibility of the existence of stable solutions
was demonstrated). In this connection, one may naively expect that the mixed Proca-Higgs systems considered here will also be unstable.
Nevertheless, to draw a definitive conclusion, the question of stability requires special studies.
Keeping in mind that the configurations considered here are described by nontopological solutions (no topological charge) and
supported by real fields (no conserved charge or particle number), there are two possible ways to study the stability.
First, one can examine the stability with respect to axisymmetric perturbations (both linear and nonlinear).
Second, it is possible to study the stability within catastrophe theory~\cite{Kusmartsev:1990cr}.
In any case, since the stability analysis requires a fair amount of effort, we plan to do this in a separate work.

\section*{Acknowledgements}

The work was supported by the Program No.~BR10965191
(Complex Research in Nuclear and Radiation Physics, High Energy Physics and Cosmology for the Development of Competitive Technologies)
of the Ministry of Education and Science of the Republic of Kazakhstan.
We are also grateful to the Research Group Linkage Programme of the Alexander von Humboldt Foundation for the support of this research.

\end{document}